\documentclass{llncs}
\usepackage{amsmath, amssymb}
\usepackage{graphicx}
\usepackage{subfigure}
\usepackage{ifthen}
\usepackage{algorithm}
\usepackage{algpseudocode}
\usepackage{enumitem}

\title{Survival Network Design of Doubling Dimension Metrics}

\author{Hao-Hsiang Hung\thanks{Math \& CS Department,
        Emory University, {\tt hhung2@mathcs.emory.edu}}
        }
\institute{Emory University}
\index{Hung, Hao-Hsiang}

\begin{document}
\thispagestyle{empty}
\maketitle

\begin{abstract}
We investigate the Minimum Weight 2-Edge-Connected Spanning Subgraph (2-ECSS) problem in an arbitrary metric space of doubling dimension and show a polynomial time randomized $(1+\epsilon)$-approximation algorithm.

\end{abstract}

\section{Introduction}
Survival network design is one of the important issues in the telecommunication field, especially when fault-tolerant is required[Vazirani steiner network]~\cite{Guptasurvey}.
Such a design assumes that it is unlikely both of them would malfunction simultaneously.
The design aims to discover reliable (more than one) paths between specific terminals in case that one of the communication path breaks down, there is still another alternative path to redirect the communication flow.

There are several variations of the problems (see metaheuristics book[book number]).
We may formalized a classical survival network design as a graph problem in the following manner:
We say a edge-weighted graph $G$ is 2-edge-connected (2-EC) if for all $e\in E$, $G-e$ is still connected.
Here we assume all the edge weights are non-negative, $w\in{\mathbb{R}}_{+}^{E}$.
The \emph{2-edge-connected-spanning-subgraph} (2-ECSS) \emph{problem} is to find a 2-ECSS of $G$ of minimum weight.
One of the reason of its proposal~\cite{Csaba02} comes from study duality of dense/sparse instances with variations of path problems related to TSP.

2-ECSS is known to be NP-hard and MAX-SNP-hard in general graphs, which is similar to metric TSP.
Note it is MAX-SNP-hard even for bounded degree graphs~\cite{Grigni07}.
In general graphs, the approximation ratio has been improved from 3 [jaja] to 2[khuller and vishkin] (the best known ratio).
About the hardness of approximation, it is NP-hard to approximate within $1573/1572$ on graphs of maximum degree 3[citation].

The problem remains NP-hard even restricted to planar graphs.
However, additive approximation algorithms have been successfully developed by Berger and Grigni~\cite{Grigni05,Grigni07}, based on the ingredient of separators in planar graphs~\cite{Grigni00}.
In particular, the time complexity of their approach is $n^{O(1/{\epsilon}^2)}$.
Borradaile and Klein developed the first EPTAS for 2-ECSS in planar graphs with running time $O(n\log n)$~\cite{Borradaile08} and has been applied to bounded genus graphs~\cite{Borradaile09}.
There are also LP-based approach to more general $k$-ECSS problem [rewrite this sentences]~\cite{Pritchard10}.


Doubling Dimension is proposed as a characteristic property to define an arbitrary metric space. [arora's lecture CS597D]
In particular, it defines the dimension from the concept of selecting balls of radius $R$ around a collection of center points to cover all the points of the given metric space $(X,d)$.
The value of dimension $D$ is impacted by the choice of $R$.

\emph{Our work}. We show the first PTAS for 2-ECSS problem in bounded doubling dimension.
Note the technique introduced by Bartel {et al.}~\cite{Bartal12} could be easily adapted to 2-ECSS problems.

\begin{theorem}
There is a polynomial time randomzied algorithm for computing the 2-ECSS problem in metric spaces of bounded doubling dimension. 
\end{theorem}

The paper is organzied as follows.
We define notations in Section~\ref{prelim}.
We show the approach for computing 2-ECSS in metric space of bounded doubling dimension in Section~\ref{method}.
In Section 4 we survey other more generalized survival network problems and discuss the feasibility of apply techniques from this paper.

\section{Preliminary}
\label{prelim}

A graph $G=(V,E)$ is 2-edge-connected if for any pair of distinct vertices $u,v\in V$, there are at least two paths connecting them.
If $G$ is edge-weighted and $w(e)>0$ for all $e\in E$, we define \textbf{2-Edge-Connected Spanning Subgraph} (2-ECSS) problem as to find a 2-ECSS of $G$ of minimum weight.

We assume the readers are familiar with the pioneering tools use in Euclidean TSP~\cite{Arora98} and the first QPTAS for TSP in doubling dimensional metric space~\cite{Talwar04}, therefore in this section we only list the important definiton and lemmata as follows without providing proofs.

\subsection{Doubling Dimension}
We say a metric space $(X,d)$ has \textbf{Doubling Dimension} $D$ if for every $S\subseteq X$, suppose all $|S|$ points could be bounded by a ball of radius $2r$, then these points could also be bounded by $2^D$ balls of radius $r$.
Note $d(x,y)$ is the distance function defined on any $x,y\in X$ with range $(0, {\mathbb{R}}_{+})$. It is symmetric, meeting the triangle inequality, and $d(x,y)=0$ when $x=y$.
For any $Y\subseteq X$, let $diam(Y)$ be the diameter and $inter(Y)$ be the minimum interpoint distance, we define its \textbf{Aspect Ratio} as $\dfrac{diam(Y)}{inter(Y)}$.
Suppose for any $x\in X$, there is a $y\in Y$ such that $d(x,y)\leq\alpha$, then we say $Y$ is a $\alpha$-covering of $X$.
If $d(y,y^\prime)\geq\beta$ for all $y,y^\prime\in Y$, then we say $Y$ is a $\beta$-packing. 
If $Y$ is $\alpha$-covering of $X$, $\beta$-packing, and $\alpha=\beta$, then we say $Y$ is a $\alpha$(or $\beta$)-net.\newline

\begin{lemma}[Packing Lemma]
Let $(X,d)$ be a doubling metric with dimension $k$, $Y\subseteq X$ with aspect ratio $\Delta$.
The $|Y|\leq (2\Delta)^k$.
\end{lemma}

\noindent\textbf{Randomized Clustering Partition}. Given a doubling metric $(X,d)$, we say $\mathbb{C}$ is a randomized clustering partition if points in $X$ are grouped into collections of subsets $\{\{S_i\}|0\leq i\leq L, L > 0, i\in I\}$ (we called them \textbf{Clusters}) which satisfy the following conditions: (1) $\forall s,t\in S_i$,$~s\cap t=\phi$; (2) $\bigcup_s s\in S_i = X$ for all $i$; (3) $S_0 = X$, $S_L$ are elements of $X$, and (4) for every $s\in S_i$, $\exists u\in s$ such that $s\subset B(u,r(i))$, where $B(u,r(i))$ is a ball with center $u$ and radius $r(i)$.
We say $s\in S_{i+1}$ is a child cluster of $t\in S_i$ if $s\subset t$.
For a cluster $s\in S_i$, we define \textbf{portals} as points in $s$ which is a set of $\alpha$-nets.


\subsection{Bound the Weight of MST in Sub-metric Space}

Let $MST(\cdot)$ be a minimum spanning tree of a (sub)-metric space and $w(\cdot)$ be the weight function of a structure.
For example, $w(MST(X))$ is the total weight of edges belong to a minimum spanning tree of a metric space $X$.

\begin{lemma}[Spanning Tree Lemma~\cite{Talwar04}]
Given a doubling metric $X$ with dimension $k$, $w(MST(X^\prime))\leq 4|X^\prime|^{1-(1/k)}\cdot diam(X^\prime)$ for all $X^\prime\subseteq X$.
\end{lemma}

As a useful side product of the diameter bound, we introduced the patching lemma 
(introduced in ~\cite{Arora98}) as an important tool for metric optimization problems in doubling metrics~\cite{Talwar04} 

\begin{corollary}[Patching Lemma]
In a doubling metric, for a tour $T$ which crosses a cluster $C$ $r$ times at crossing points $P$, then there is another tour $T^\prime$ such that $T^\prime$ crosses $C$ at most twice and $w(T^\prime)\leq w(T)+4w(MST(P))$.
\end{corollary}

Note the patching lemma is an application of the spanning tree lemma. 
The following simple extension shows that given a tour $T$ which visits a cluster $C$ contained in a ball $B$, we can isolate another tour connecting the crossing points of $C$ and segments outside of $B$.

\begin{corollary}[Clustering Tour Lemma]
Suppose $(X,d)$ is a doubling dimension, and $C\subset X$ is a cluster contained in a ball $B$, and $T$ is a tour crosses $C$ at points $P$.
Let $\{T_i|T_i\subseteq T\}$ such that edges in $T_i$ has at most one end inside of $B$.
Then there is another tour $T^\prime=\{T_i\}\bigcup MST(P)$ such that $w(T^\prime)\leq 16r^{1-(1/k)}diam(B)+\sum_{i=1}^k w(T_i)$.
\end{corollary}

Additionally, given a doubling metric $(X,d)$ with a randomized clustering partition $\pi$, and a tour $T$ of $(X,d)$, we can convert $(X,d)$ into another tour $T^\prime$ called \textbf{well-behaved tour} with total weights bounded by the weight of $T$, and $T^\prime$ only enters/leaves the clusters via portals.

\begin{lemma}[Well-behaved Tour~\cite{Bartal12}]
$w(T^\prime)\leq (1+6\epsilon)w(T)$.
\end{lemma}

\begin{proof}
For any edge $e=(x,y)$ which is cut by a ball of a cluster in $\pi$, starting from $x\in S_L$, we go upward until we reach $x^\prime\in S_i$ such that $s^i\leq\epsilon d(x,y)$, reaching some $y^\prime\in S_i$, and go downwards to reach $y\in S_L$.
Now the detour path becomes $x\cdots x^\prime y^\prime\cdots y$.
By geometric sum we know $w(x\cdots x^\prime)\leq 2\epsilon d(x,y)$ (so is $y^\prime\cdots y$).
Besides, $d(x^\prime, y^\prime)\leq(1+2\epsilon)d(x,y)$ by the triangle inequality.
\qed
\end{proof}

\section{2-ECSS in Doubling Dimension}
\label{method}

As the classical approach for metrical optimization problems in doubling dimension setting, our ingredients are from ~\cite{Arora98}~\cite{Talwar04} and we also adapt the concept from ~\cite{Bartal12} for TSP in metric space of doubling dimension.
The key step of improvement from QPTAS to PTAS is the parameterized sparsity concept called $q$-sparse (will be mentioned later), which bounds the crossing number to $O(\log n)^c$ (here $c<1$).
However, although it is a PTAS, the exponent might still be too large.

We sketch the high level algorithmic steps as follows.
\begin{algorithm}[htb]
  \caption{The approximation algorithm for 2-ECSS in doubling dimension}
  \label{alg:Framework}
  \begin{algorithmic}[1]
    \Require
      A metric space $(X,d)$ with doubling dimension $D$ and aspect ratio $R$
    \State Randomized decompose $(X,d)$ into a tree of level $\log R$,
    \State with internal node (balls of radius $2^i$) at level $i$
    \State Set the number for sparsity $q$ with value ${(\dfrac{s}{\epsilon})}^{O(k)}\cdot2^{O(k^2)}$ and let $q^\prime=O(q\sqrt[8]{\log n})$
    \If{$(X,d)$ is not $q^\prime$-sparse}
    \State Split the points into dense subsets $X_d$ and sparse subsets $X_s$
    \State Recursively solve sub metric space $(X_d, d)$ with aspect ratio $R_d$
    \State Combine best local tours from $X_d$ and $X_s$
    \EndIf
    \State In each sparse set:
    \State Choose the proper number limit $m$ of nets for each cluster
    \State Choose the proper crossing number limit $r$ for each cluster
    \For{$i$ from $1$ to $\log R$}
    \For{Each cluster of level $i$ with portals}
    \State Retrieve local optimal tours $\{T_i\}$ from children clusters (of level $i-1$)
    \State Use Patching Lemma to reduce $k$ to 2
    \State Use DP to calculate the local optimal tour of 2-ECSS by MST
    \EndFor
    \EndFor
  \end{algorithmic}
\end{algorithm}

\subsection{Randomized Clustering Partition}
We use a metric space $(X,d)$ of doubling dimension into smaller level of clusters.
The first level include all the points in $X$ (we assume the aspect ratio $R$ is known and the minimum distance is 1.
Generally, for the points of each cluster in the $i$-th level, the dimension bound guarantees that there are $2^k$ balls of radius $\frac{R}{2^{i-2}}$ of clusters of $(i+1)$-th level whose union could cover all these points.
In this part the algorithm works in a randomized divide and conquer manner.
We provide a high-level of sketch of it as follows.
Assume that every vertex has its own index; at each level $i$, we just generate a random permuation $\pi$ of $[n]$ and follow the ordering of $\pi$ to select centers for the balls (of diameter $\frac{R}{2^{i-1}}$).
A point is assigned to the first ball that covers it.

The following lemata correspond to in~\cite{Talwar04} and~\cite{Bartal12}. 

\begin{lemma}[Cut Property]
Let $u,v\in X$ with distance $d(u,v)$ in the doubling metric $(X,d)$ with dimension $k$. 
The probability that $u$ and $v$ are divided into different clusters (of the same level $i$) is at most $O(k)\frac{d(u,v)}{2^i}$.
\end{lemma}

\subsection{Sparse Subspace}
\begin{definition}[Sparsity]
We say a tour $T$ is $q$-sparse if for each point $x\in X$ and for all level of clusters in which $x$ belongs, the total weight $w(T\bigcap B^\star(u,3s^i))\leq q s^i$. 
\end{definition}


We first show that there is a PTAS for TSP if a tour $T$ in a metric space $(X,d)$ is $q$-sparse. 
If not then $(X,d)$ can be decomposed into two subspaces $\{X_1, X_2\}$ where $X_1$ is $q^\prime$-sparse and the cost for combing the optimal well-behaved subtours is not too high.

\textbf{Choice of Radius}. The range of the radius for any portal $p$ of level $i$ is $p(i)=[s^i, 2s^i]$.
Suppose we have an optimal well-behaved tour $T_{opt}$.
Classify edges of $T_{opt}$ into \textbf{short edges} (that is, edges of length at most $s_i$) and \textbf{long edges}.
We want to choose the proper range $p(i)$ such that the number of short edges of $T$ cut by the ball $B(p, p(i))$ is fewer than $10qk$.
If there is a decomposition scheme to choose portals in every level so that short edges are cut with high probability, we will have the following good property.

\begin{lemma}[Sparse Lemma]
Given a doubling dimensional metric space $(X,d)$ which admits a well-behaved tour $q$-sparse tour $T$, then there is a randomized polynomial time algorithm for finding another $T^\prime$ such that $w(T^\prime)\leq(1+\epsilon)w(T)$ with constant probability.
\end{lemma}
\begin{proof}
First we bound the total weight of short edges of $T$ inside $B(p,3s^i)$, which is at most $qs^i$ by the assumption that $T$ is $q$-sparse, and the same bound applies to short edges with one end inside $B(p,2s^i)$.
Let $V$ be the set of values in $[s^i,2s^i]$ such that fewer than $10qk$ short edges are cut by the ball $B(p, p(i))$, and if the distribution of edges inside the ball is uniform, then at least $1-\frac{1}{10k}$ of them belong to $V$.

For finding a proper value in $V$ we might need resampling to fit the sparsity assumption.
The radius is chosen from a exponential distribution with density function $\frac{2^{8k}}{1-2^{-8k}}\cdot\frac{8\ln 2 k}{s^i}\cdot2^{\frac{8k}{s^i}r}$.
By using the Padding Lemma in~\cite{abn11}, the sampled radius belong to $V$ is at least $1/2$. (illustrating the worst case scenario).

We then bound the cost of converting $T$ to another $T^\prime$ such that the size bound of the portals (of every level) and the allowable visiting times of portals are bounded, based on building a randomized decomposition of the metric space with portals specified for each level.
Note from lemma [ClusteringCut] the probability that an edge $(u,v)$ is cut by a ball of level $i$ is $\frac{2c^\prime d(u,v)k}{s^i}$, and all of them needs to find a $\frac{s^i}{M}$-net point, introducing additional length of at most $\frac{2s^i}{M}$.
Therefore for one particular $e$ to be cut at level is $O(\frac{\epsilon d(u,v)}{\log_s n})$, and by summing all the levels we get $O(\epsilon d(u,v))$.

Next we bound $r$, the number of entering and leaving of edges in $T^\prime$ through portals in the decomposition tree of clusters based above. Again, two cases are discussed separately.
For \textbf{short} edges, we know that the radius range of cluster at level $i$ is $[s^i, 2s^i]$, and by the packing lemma, a ball of level $i$ centered at $u$ may neighbor to at most $2^{4k}$ neighbors balls at least at level $i$ (and all level $i > j$), and if we overcount each of them might provide $10qk$ edges to be cut by the ball centered at $u$.
Now we set up the upper bound $r^\prime$ for the number of short edges $r(short)$ as $\max\{2^{4k}\cdot 10qk\cdot\log_s{(\log n)}, (\frac{2c^\prime k}{\epsilon})^{k}\}$, where the $\log_s(\log n)$ counts for the allowable levels need not patching (by going through portals).
Note such value of $r^\prime$ guarantees that for every level $i\geq j+\log_s(\log n)$, there are at least $r^\prime/2$ edges must be cut by the ball (of level $i$).
If unfortunately more than $r^\prime$ edges are cut by $C$ then we have to build the spanning tree between portals and then apply patching lemma, and the cost shared by edge is $O(\frac{s^j r^{\prime 1-\frac{1}{k}}}{r^\prime/2})$.
Multiply by the probability that an edge at level $j$ needs patching is $O(\frac{d(u,v)k}{s^{j+\log_s(\log n)}})$ ($j$ can only be larger), and multiply by all possible $\log n$ levels $(u,v)$ might be cut, the patching cost is $O(\epsilon d(u,v))$.

On the other hand, for handling \textbf{long} edges we need $r^{\prime\prime}=(s/\epsilon)^{2k}$ crossings at portals.
We directly count their ends inside of the ball as the portals.
We analyze it as follows.
Instead of considering a ball of center $u$ with radius $s^j$, we now consider a collection of eccentric balls of different radius: $s^j$, $3s^j$, and $4s^j$ and called these balls $B_1$, $B_3$, and $B_4$.
Since $B_3$ contains $B_1$, we use the number of long edges $B_3$ to bound that of $B_1$.
The maximum number of possible distinct $s^j$-net points inside $B_3$ is at most $(\frac{3s^j}{\epsilon s^j/s})^k\leq \frac{1}{2}(s/\epsilon)^{2k}$, and we may assume that by patching lemma every $s^j$-net points inside $B_3$ is visited at most twice.
Connecting all these edges inwards until reach the interior of $B_1$, we can use them to bound the number of long edges of $B_1$, that is, $(s/\epsilon)^{2k}$.

Summing the required crossing in short and long edges, $r=r^\prime+r^{\prime\prime}=2^{4k}\cdot10qk\log_s{\log n}+(2c^\prime k/\epsilon)^k+(s/\epsilon)^{2k}$ is a rough upper bound of the crossing number required.
\qed
\end{proof}

\subsection{Algorithms for 2ECSS in a Sparse Doubling Metric}

Now we provide a PTAS for 2-ECSS in doubling metrics.
For proper calculation purpose, we set $s=(\log n)^{1/(ck)}$ for some $c\geq 32$ in the following algorithm.
\begin{theorem}\label{longshort}
Given a doubling metric $(X,d)$ which admits a $q$-sparse well-behaved tour $T$, there is a polynomial time randomized algorithm to find a tour $T^\prime$ such that $w(T^\prime)\leq(1+\epsilon)w(T)$.
\end{theorem}

\begin{proof}
By the small fractional "bad events" in the sparse lemma and the exponential distribution assumption we know a successful "good" random guess for the radius is $1/2$ (for balls of any level), which means we might need $O(\log n)$ times of independent guessing so that the probability of one of the successful choice is $1-\frac{1}{n^2}$. We apply the method with union bound to determine the raidus of each net point.

We next determine the level of dynamic programming.
By applying the above choice of $s$ and $c$, the number of levels for the clusters is $L=O(\log_s^n)=O(\frac{k\log n}{\log\log n})$.

For combing the partial solutions from child clusters of level $i$ to one cluster $C$ of level $i-1$, first we have to count the $O(\log n)$ random choices for the radius within range $[s^{(i-1)},2s^{(i-1)}]$.
The number of other net points of the same level ($i-1$) whose radii might also cut by $C$ is at most $2^{4k}$, therefore the possible configuration of guessing radius and cutting with neighboring net points is $(O(\log n))^{2^{4k}}$.
Roughly multiply $L$ (for every level it is counted) the possible formation of $C$ is $(O(\log n))^{2^{4kL}}=n^{O(2^{5k})}$.

We now consider another factor: the well-behaved tours. 
Because $C$ has at most $s^{2k}$ child clusters, there are at most ${O(\log n)}^{2^{4k}{s^{2k}}}$ possible parent-child configurations.
We do not have to consider the radii of lower level balls into cut in the current level because they are they are fixed.
The maximum access time to partial solutions from children clusters is $(m^r)^{s^{2k}}$ possibilities.

For the cost from edges coming from the $i$-th level children clusters to the $(i-1)$-the level cluster, we upper bound it by pairing up all the possible combination: $(rs^{2k})^{rs^{2k}}$.

The total complexity is a huge constant number, multiply by a huge polynomial:
They are: $O(mr\log n)^{rs^{2k}}=2^{O(s\cdot q\log q(k/\epsilon)^{4k}(\log\log n)^2)}$ (interacting with children clusters) and $n^{O(2^{5k})}$ (considering cuts between higher levels).
\end{proof}

\subsection{Decomposing a Doubling Metric into Sparse and Dense Parts}
\begin{lemma}[Sparse Decomposition] 
Given a doubling metric $(X,d)$, and let $OPT(X^\prime)$ be the optimal well-behaved tour for $X^\prime\subseteq X$.
There is a polynomial time randomized algorithm to decompose $X$ into two parts $X_1$ and $X_2$ such that
\begin{enumerate}
\item $X_1\subset X$, $X_2\subset X$, $X_1\bigcup X_2=X$, and $X_1\bigcap X_2=\phi$;
\item The optimal well-behaved tour of $X_1$ is $q^\prime$-sparse ($q^\prime=O(q\sqrt[8]{\log n})$);
\item $w(OPT(X_1))+w(OPT(X_2))\leq w(OPT(X))+\epsilon w(OPT(X_1))$. 
\end{enumerate}
\end{lemma}
\begin{proof}
For calculation purpose, we set the sparsity parameter $q$ be $(s/\epsilon)^{O(k)}\cdot 2^{O(k^2)}$.
The sparse testing starts from level $L$ upwards until we reach some cluster in level $i$ which is not $q$-sparse, and we pick the densest cluster (say, with center $v$) in this level.
For all $v$ in all levels $i$ we examine if $w(MST(v,3s^i)) > 2qs^i$.
If no such $v$ exists then we are done, because by the Local Tour Lemma [to be added later] we know there is a $q^\prime=13q$ which satisfies property 2.
Note here we let $X_2$ be an arbitrary point, and $X_1=X$, which meets property 1.
Because it is unnecessary to connect $X_2$ to $X_1$, property 3 definitely holds.

Consider the case when there is such a center point $v$ of a cluster in level $i$.
Let $q^\star=w(MST(B(v,3s^i)))/(2s^i)$.
Note $q^\star>q$ always holds (independent of the value of $q$).
If we partition $X$ into two disjoint subsets, the total weight of the crossing edges (subtours) may be uncontrollable, we need to bound the total length of the crossing edges via the similar idea used in Theorem~\ref{longshort}: we classify them into long and short edges, by scanning another eccentric ball of larger radius and bound of cost of these crossings.
Suppose $X_1$ is the sparse part that we found, and $X_2$ is the dense part.
Now $X_2$ needs to be recursively partitioned (because it is still dense and we subtitute $X_2$ with $X$ with sparsity checking).
The original ball has radius $3s^i$, and we may pick up a value $h$ in $[12s^i, 13s^i]$(we will prove the existence of $h$ later and how to choose the best value).
A slight difference here is that we define the long crossing edges of the tour $T$ in the layer between $B(v,3s^i)$ and $B(v,h)$, with length more than $\delta s^i$ such that $\delta=O(\epsilon/2^{10k})$.
For patching into a well-behaved tour $T^\prime$, we consider $T^\prime$ traverses upwards until level $l$ such that $s^l\leq \epsilon\delta s^i$.
Note that we assume $\epsilon<\frac{1}{72}$ and set $\delta=O(\epsilon/2^{10k})$ with a large constant.
After reaching level $l$ we apply Well-behaved Tour lemma, and then we connect subtours between portals by Clustering Tour Lemma.
We multiply the upper bound of the Clustering Tour Lemma $(1+6\epsilon)$ by the maximum possible number of portals at level $l$ ($(s/\epsilon\delta)^{4k}$) and the worst possible tour between every pair of portal $O(s^i)$.
The result shows patching these long edges needs additional cost
\begin{equation}
\epsilon q s^i. 
\end{equation}
On the other hand, each short edge has one end with distance $h+\delta s^i$ from $v$.
We patch them by connecting them to portals of level $l^\prime$ such that $s^{l^\prime}\leq\delta s^i$.
Therefore we need to include portals (*portal net points interchangeably check) of level $l^\prime$ from $X_2$.
Once we obtain the temporary sub-tour for portals of level $l^\prime$, the additional patch could be done by applying the Well-behaved Tour Lemma.

To bound the patching cost for short edges, we count the number of $l^\prime$-level portals between the chosen radius with range $(h-\delta s^i, h+\delta s^i)$.
Here the principle is to choose $h$ such that the total weight of the collection of MST with radius $4l^\prime$ is minimized (to bound the cost of patching by converting the original tour to a well-behaved one).
The maximum possible number of $l^\prime$-level portals is bounded by $(2\cdot 14s/\delta)^k\leq (s/\delta)^{2k}$.
By a similar analysis in the Sparse Lemma, for each $l^\prime$-level portals, we can bound the total cost of patching for short edges (they all have to modify to leave the ball from these portals) with $w(MST(B(u,s^{l^\prime})))\leq w(T\bigcap B^\star(u,4s^{l^\prime}))$.
If we choose the best $h$ so that the total weight of the MST with radius $4s^{l^\prime}$ is minimized, we obtain
the following: 
\begin{equation}
\sum_{u\in N(h)}w(MST(B(u,l^\prime)))\leq\sum_{u\in N(h)}[w(T\bigcap B^\star(u,4s^{l^\prime}))]+(s^4/{\epsilon^2\delta^2})^ks^i
\end{equation}
Note we denote $N(h)$ as the collection of $l^\prime$ portals in the layers between two eccentric balls of radii $h-\delta s^i$ and $h+\delta s^i)$ respectively. 
In addition, if we denote $A^\star(v,r_1,r_2)$ means the number of edges inside of the layers of eccentric balls between radius $r_1$ and $r_2$, then we know $\sum_{u\in N(h)}[w(T\bigcap B^\star(u,4s^{l^\prime}))]\leq 2^{5k}w(T\bigcap A^\star(v,h-5\delta s^i,h+5\delta s^i))$, because each ball $B(u,s^{l^\prime})$ centered from every $u\in N(h)$ intersects at most $2^{5k}$ other balls centered at portals of $l^\prime$ level: $(\frac{2*2*4(s^{l^\prime})}{s^{l^\prime-1}})^k=O(2^{5k})$ (suppose $s=4$).

Because our goal is to minimize $\sum_{u\in N(h)}w(MST(B(u,l^\prime)))$ by choosing the proper $h$, we want to know how heavy the best setting of $h$ could still weigh.
Recall that we assume the $q$-sparse test fails (of radius $3s^i$ at some point maximizing $w(MST(B(v,3s^i)))$), then we enlarge the radius to $13s^i$ to so that we can bound the total weight of the MST inside the ball of radius $3s^i$ by the sub-paths of a well-behaved tour inside of the ball of $13s^i$, plus the weight of some long edges.
By the packing property, it turns out that $B(v,13s^i)$ can be covered by at most $2^{4k}$ balls of radius $3s^i$, and the patching cost of connecting the centers of these balls together is at most $2\cdot13s^i$, so $w(MST(B(v,13s^i)))\leq 2^{4k}w(MST(B(v,3s^i)))+2^{4k}26s^i\leq 2^{5k}\cdot 3q^{\star}s^i$, where $q^\star=w(MST(B(v,3s^i)))/(2s^i) > q$ because $v$ is the worst point which fails the sparsity test in $X$($v$ is also in $X_1$).
Therefore we obtain $\sum_{u\in N(h)}w(T\bigcap B^\star(u,4s^{l^\prime}))\leq O(\delta)\cdot 2^{10k}q^\star s^i$ which bounds the patching cost for edges between the extension of $X_1$ (including some points from $X_2$) and the extension of $X_2$, and since we set $\delta=O(\epsilon/2^{10k})$, by substitution we get
\begin{equation}
\epsilon q^\star s^i.
\end{equation}
Summing the above equations, the total cost of patching is bounded by 
\begin{equation}
O(\epsilon)\cdot w(OPT(X_1)),
\end{equation}
where $w(OPT(X_1))$ is the total weight of a well-behaved tour in $X_1$.

We apply similar patching and analysis to the extension of $X_2$ (we also have to add portals of level $l$ for long crossing edges and of $l^\prime$ for short crossing edges), and once we know the partial subtours we simply join them together (because they overlap).

To prove the well-behaved tour of $X_1$ (union its extension) is $q^\prime$-sparse where $q^\prime=O(q\sqrt[8]{\log n})$, first note that we wanted to find a $v$ of level $i$ such that $w(MST(v,3s^i)) > 2qs^i$.
Before we reached level $i$ (assume it exists), for each ball of each previous lower level $j$ ($j<i$ and $s^j\leq s^i$), $w(MST(u,3s^j))\leq2qs^j$ (for some $u$).
By a common-known fact that a tour of points in $X_1$ is bounded by $2w(MST(X_1))$, combined with the Well-behaved Tour Lemma, we know a well-behaved tour of $X_1$ inside of $B^\star(u,3s^j)$ (here by $B^\star$ we mean all pairwise paths of points in the ball) is bounded by $13qs^j$.
Summing all $j$ we bound the new parameter for sparsity.

Last we show property 3 holds.
The union of two well-behaved subtours $OPT(X_1)$ and $OPT(X_2)$ is definitely a well-behaved tour for $X$ because they both visited portals in level $l^\prime$.
The only additional cost comes from counting the crossing edges from $X_1$ to $X_2$ by taking the paths through the level-$l^\prime$ portals, which is at most $\epsilon w(OPT(X_1))$ by equation (4).
\end{proof}

\begin{lemma}
Suppose the well-behaved subtours of $X_1$ and of $X_2$ are known and $OPT(S)$ is the optimal well-behaved tour for TSP in the subset of points $S\subseteq X$, and $T(S)$ is the solution tour by the above computation.
Then $w(T(X))\leq (1+O(\epsilon))\cdot w(OPT(X))$.
\end{lemma}
\begin{proof}
We first prove $w(T(X))\leq(\dfrac{1+\epsilon}{1-\epsilon})\cdot w(OPT(X))$, then we apply the Well-behaved Tour Lemma.
In the trivial case $X$ contains one point so no decomposition is required.
By induction we suppose $w(T(X_2))\leq(\frac{1+\epsilon}{1-\epsilon})\cdot w(OPT(X_2))$.
By the Sparse Decomposition Lemma we know 
\begin{equation}
w(T(X_1))\leq (1+\epsilon)w(OPT(X_1))
\leq (\dfrac{1+\epsilon}{1-\epsilon})\cdot (w(OPT(X)-w(OPT(X_2))))
\end{equation}
Combined with the hypothesis, we get $w(T(X))\leq(\dfrac{1+\epsilon}{1-\epsilon})\cdot w(OPT(X))$; then we apply the Well-behaved Tour Lemma to obtain $w(T(X))\leq (1+O(\epsilon))\cdot w(OPT(X))$, where $\epsilon$ is adjustable to get the target approximation ratio.
\end{proof}


\section{Discussion}

Because of the wide applications of the portal framework, the approach of Bartel {et al.}~\cite{Bartal12} in solving metric TSP in doubling metrics could be easily adapted for some classical survivable network design problems.
Note that a question asked by Berger and Grigni~\cite{Grigni07} about algorithms for Steiner version of 2-ECSS in planar graphs has been answered~\cite{Borradaile08} and even extended to bounded genus case~\cite{Borradaile09}.
We are interested if the Steiner 2-ECSS could be found in doubling metrics.

Another natural generalization is for $k$-connectivity, which has been studied well in geometric graphs~\cite{Czumaj06}.
Note in their patching lemma, they reduce the number of crossing by using Steiner points, based on the assumptions that they could remove the ends of crossing edges and replace them with collection of points and they assume the cost of edges between these points are zero(or infinitesimally small).
Such a scheme might not well fit into the doubling metrics and new technique is needed.

\bibliographystyle{abbrv}
\bibliography{2ecss}

\begin{thebibliography}{10}

\bibitem{abn11}
I.~Abraham, Y.~Bartal, and O.~Neiman.
\newblock {Advances in metric embedding theory}.
\newblock {\em Advances in Mathematics}, 228:3026--3126, 2011.

\bibitem{Arora98}
S.~Arora.
\newblock Polynomial time approximation schemes for euclidean traveling
  salesman and other geometric problems.
\newblock {\em J. ACM}, 45(5):753--782, Sept. 1998.

\bibitem{Bartal12}
Y.~Bartal, L.-A. Gottlieb, and R.~Krauthgamer.
\newblock The traveling salesman problem: low-dimensionality implies a
  polynomial time approximation scheme.
\newblock In {\em Proceedings of the 44th symposium on Theory of Computing},
  STOC '12, pages 663--672, New York, NY, USA, 2012. ACM.

\bibitem{Grigni05}
A.~Berger, A.~Czumaj, M.~Grigni, and H.~Zhao.
\newblock Approximation schemes for minimum 2-connected spanning subgraphs in
  weighted planar graphs.
\newblock In {\em Proceedings of the 13th annual European conference on
  Algorithms}, ESA'05, pages 472--483, Berlin, Heidelberg, 2005.
  Springer-Verlag.

\bibitem{Grigni07}
A.~Berger and M.~Grigni.
\newblock Minimum weight 2-edge-connected spanning subgraphs in planar graphs.
\newblock In L.~Arge, C.~Cachin, T.~Jurdzinski, and A.~Tarlecki, editors, {\em
  ICALP}, volume 4596 of {\em Lecture Notes in Computer Science}, pages
  90--101. Springer, 2007.

\bibitem{Borradaile08}
G.~Borradaile and P.~Klein.
\newblock The two-edge connectivity survivable network problem in planar
  graphs.
\newblock In {\em Proceedings of the 35th international colloquium on Automata,
  Languages and Programming, Part I}, ICALP '08, pages 485--501, Berlin,
  Heidelberg, 2008. Springer-Verlag.

\bibitem{Borradaile09}
G.~Borradaile, P.~Klein, and C.~Mathieu.
\newblock An o(n log n) approximation scheme for steiner tree in planar graphs.
\newblock {\em ACM Trans. Algorithms}, 5(3):31:1--31:31, July 2009.

\bibitem{Csaba02}
B.~Csaba, M.~Karpinski, and P.~Krysta.
\newblock Approximability of dense and sparse instances of minimum
  2-connectivity, tsp and path problems.
\newblock In {\em Proceedings of the thirteenth annual ACM-SIAM symposium on
  Discrete algorithms}, SODA '02, pages 74--83, Philadelphia, PA, USA, 2002.
  Society for Industrial and Applied Mathematics.

\bibitem{Czumaj06}
A.~Czumaj and A.~Lingas.
\newblock Approximation schemes for minimum-cost k-connectivity problems in
  geometric graphs.
\newblock In T.~Gonzalez, editor, {\em Handbook of Approximation Algorithms and
  Metaheuristics}, pages 51--1~51--21. CRC Press, 2007.

\bibitem{Grigni00}
M.~Grigni and A.~P. Woloszyn.
\newblock Well-connected separators for planar graphs, 2000.

\bibitem{Guptasurvey}
A.~Gupta and J.~K\"{o}nemann.
\newblock Approximation algorithms for network design: A survey.
\newblock {\em Surveys in Operations Research and Management Science}, 16(1):3
  -- 20, 2011.

\bibitem{Pritchard10}
D.~Pritchard.
\newblock k-edge-connectivity: approximation and lp relaxation.
\newblock In {\em Proceedings of the 8th international conference on
  Approximation and online algorithms}, WAOA'10, pages 225--236, Berlin,
  Heidelberg, 2011. Springer-Verlag.

\bibitem{Talwar04}
K.~Talwar.
\newblock Bypassing the embedding: algorithms for low dimensional metrics.
\newblock In {\em Proceedings of the thirty-sixth annual ACM symposium on
  Theory of computing}, STOC '04, pages 281--290, New York, NY, USA, 2004. ACM.

\end{thebibliography}



\end{document}